\documentstyle[12pt]{article}
\begin{document}

\baselineskip 24pt

\newcommand{\sheptitle}
{Anomalous 4-Jet Events at LEP: a Signal of Low Scale Technicolour?}

\newcommand{\shepauthor}
{S. F. King,}

\newcommand{\shepaddress}
{Physics Department,\\University of Southampton,\\Southampton,\\SO9 5NH,\\U.K.}

\newcommand{\shepabstract}
{We suggest that the anomalous 4-jet events recently reported by ALEPH 
may be the first indication of low scale technicolour.
According to our interpretation about half of
the events are due to resonantly enhanced pair production of charged
technipions, of mass around 55 GeV, each decaying
into a pair of jets, the other half being due to standard processes. 
The resonant enhancement is due to 
a nearby technirho resonance in the mass range 150-200 GeV,
leading to hugely enhanced signals at the forthcoming 
LEP run at $\sqrt{s}=160$ GeV since the CM energy
would be closer to the peak of the resonance.}

\begin{titlepage}
\begin{flushright}
SHEP 96/09\\
hep-ph/9604399
\end{flushright}
\vspace{.4in}
\begin{center}
{\Huge{\bf \sheptitle}}
\bigskip \\ \shepauthor \\ {\it \shepaddress} \\ \vspace{.5in}
{\bf Abstract} \bigskip \end{center} \setcounter{page}{0}
\shepabstract
\end{titlepage}

\newcommand{\beq}{\begin{equation}}
\newcommand{\eeq}{\end{equation}}
\newcommand{\beqarr}{\begin{eqnarray}}
\newcommand{\eeqarr}{\end{eqnarray}}
\newcommand{\psibarpsi}{\mbox{$<\bar{\psi}\psi>_{M_{ETC}}$}}
\newcommand{\alphaetc}{\mbox{$\alpha_{ETC}$}}

\newpage

\setcounter{page}{1}
\pagestyle{plain}

Recently ALEPH has released a preprint \cite{ALEPH}
in which they discuss some unusual 4-jet events
which were previously reported in preliminary form.
These data are from the LEP runs consisting of
$2.8pb^{-1}$ at $\sqrt{s}\approx 130$ GeV and
$2.8pb^{-1}$ at $\sqrt{s}\approx 136$ GeV
(henceforth called LEP1.3).
ALEPH studied the dijet invariant masses of 4-jet
final states, and found an excess of 8 events with the sum of the
two dijet masses peaked at $105$ GeV.
The other LEP groups do not see such a large 4-jet enhancement
and it has been estimated that the probability that the 4-jet data
is consistent amongst all 4 groups is about $5\%$.
However details of the ALEPH events are not consistent with QCD background, 
and one possibility is the pair production of new particles \cite{ALEPH}.

The angular distribution of the
events is consistent with a roughly equal mixture of standard
processes and scalar particle production
\footnote{Neither is compatible with the angular distribution
by itself, but a cocktail
of 1/3 standard processes plus 2/3 scalar particle production
is consistent with a $25\%$ statistical error in these proportions.}
(assuming the scalar production cross-section is large enough)
\cite{ALEPH}. Under the hypothesis that roughly $50\%$ of the
events are due to standard processes, the di-jet mass difference
is consistent with either a pair of colour singlet
scalar particles or a pair of coloured scalars \cite{ALEPH}.
\footnote{Again some fraction of the events must be due to standard
processes to achieve compatibility with the di-jet mass difference. 
In the case of equal mass colour
singlets $70\%$ of the events are required to be 
due to standard processes with a large error in excess of $25\%$.
For coloured scalars or 10 GeV mass split colour singlets
the fraction of standard processes required is $30\%$ or $25\%$
respectively with similar errors.}
Studies on electric charges further imply that the production of
charged scalars is favoured over neutral scalars \cite{ALEPH}.
If the entire excess of events corresponds to
the production of a pair of new particles each with a
mass of about 55 GeV, and each decaying into a pair of jets
then it would correspond to a production
cross-section of $3.1\pm 1.7$pb, however as we have seen 
it is more likely that the new physics cross-section is
about half this value. Since no events
contain $\bar{b}b\bar{b}b$, the effect cannot be due to
$hA$ production, since both the CP-even Higgs $h$
and the CP-odd Higgs $A$ are expected to decay into
$\bar{b}b$, and in any case the expected cross-section 
would only be about 0.5 pb and neutral scalars are not favoured.

Are the events consistent with charged Higgs $H^+H^-$
production? As stated, charged scalars are preferred over neutral scalars,
and in this case the 4-jet events are consistent with
the decays $H^+\rightarrow c\bar{s}$,$H^-\rightarrow \bar{c}s$,
which are one of the common decay modes of charged Higgs bosons,
although no strangeness enhancement due to $K_S$ production
is seen.
\footnote{Only 3 $K_S$ mesons are seen compared to
$2.7\pm 1.4$ expected from a normal flavour mix and
$5.3\pm 1.6$ from $c\bar{s}\bar{c}s$ final states \cite{ALEPH}.
A better fit is achieved by assuming half the events
are due to scalar particles which decay into $c\bar{s}\bar{c}s$.}
The other decay mode of charged Higgs is
$H^+\rightarrow \tau^+\nu_{\tau}$,
$H^-\rightarrow \tau^-\bar{\nu_{\tau}}$
which is not observed. For a tau branching fraction in excess of $50\%$, 
the $95\%$ confidence
level upper limit is 1.2 $c\bar{s}\bar{c}s$ events \cite{ALEPH},
so the absence of such decays seems to count
against the charged Higgs hypothesis,
as does the expected smallness of the cross-section of about 0.4 pb for
55 GeV equal masses. Assuming that roughly 
a half of the excess of events is due
to charged Higgs the theoretical cross-section still looks too small
compared to the measured cross-section.

In this paper we shall explore the possibility that
ALEPH are pair producing equal mass charged technipions
of mass around 55 GeV
which decay predominantly into $c\bar{s}\bar{c}s$
4-jet final states. We shall assume that roughly one half of the
ALEPH 4-jet excess is due to this process, and that the
other half is due to standard processes (QCD).
The production rate will be enhanced
by nearby technirho resonances in the mass range 150-200 GeV.
In conventional technicolour (TC)
the TC confinement scale
${\Lambda}_{TC} \sim 500$ GeV so that the
technirho has a mass of about 1 TeV \cite{tc} which is too high for the 
purpose of enhancing the technipion production rate.
However recently we discussed the phenomenology of an
$SU(2)_{TC}$ technicolour model
with a low technicolour confinement scale 
${\Lambda}_{TC} \sim 50-100$ GeV \cite{lstc}
\footnote{We define ${\Lambda}_{TC}$ to be equal to 
half the mass of the lowest lying vector resonance.} 
and discussed charged technipion production, 
with the production cross-section resonantly enhanced
by the presence of a light technirho resonance \cite{lstc}.
We now return to this idea in the light of the ALEPH 4-jet excess.
As we shall see, assuming that ALEPH are seeing such a signal,
the implications of this scenario
for the LEP run in June of this year at $\sqrt{s}=160$ GeV
(henceforth called LEP 1.6) are so great that it as well
that our experimental and phenomenological colleagues are 
made aware of them now. This is our principal motivation
for writing this paper at the present time.

The basic production mechanism we suggest is:
\beq
e^+e^-\rightarrow \gamma^{\ast},Z^{\ast}\rightarrow 
\rho_{TC}^{\ast} \rightarrow \pi_{TC}^{+}\pi_{TC}^{-}
\label{mechanism}
\eeq
The partial width of the technirho 
\footnote{In fact there may be more than one technirho, however
for simplicity we shall assume a single technirho resonance.}
into a single pair of charged technipions is given by a scaling argument as,
\beq
\Gamma (\rho_{TC}\rightarrow \pi^{+}_{TC} \pi^{-}_{TC})
\approx \frac{m_{\rho_{TC}}}{m_{\rho}}
\frac{\beta_{\pi_{TC}}^3}{\beta_{\pi}^3}\Gamma (\rho \rightarrow \pi^+ \pi^-)
\label{width}
\eeq
where $\beta_{\pi (TC)}=(1-\frac{4m_{\pi (TC) }^2}{m_{\rho (TC) }^2})^{1/2}$
is the relevant phase space factor where $m_{\pi_{TC}}$ is the 
technipion mass.
In this case the full width of a technirho is approximately
equal to the partial width into technipions,
\beq
\Gamma_{\rho_{TC}}\approx
\Gamma (\rho_{TC}\rightarrow \pi^{+}_{TC} \pi^{-}_{TC}).
\eeq
It is clear that the presence of a light technirho will serve
to resonantly enhance the production of charged technipions
relative to that of charged Higgs bosons of the same mass.
According to VMD arguments (see later), in the resonance region
we expect this enhancement to be given by a factor $R$ where
\beq
R=\frac{\sigma(e^+e^-\rightarrow \pi_{TC}^{+}\pi_{TC}^{-})}
     {\sigma(e^+e^-\rightarrow H^+H^-)}
\approx        \frac{m_{{\rho}_{TC}}^4}
{\left[(s-m_{{\rho}_{TC}}^2)^2+
\Gamma_{\rho_{TC}}^2m_{{\rho}_{TC}}^2\right]}
\label{ratio}
\eeq
where $\sigma(e^+e^-\rightarrow H^+H^-)$ is the usual production
cross-section for charged Higgs bosons in a two-Higgs doublet model.
This expression is valid for any charged technipion mass (assumed to be 
approximately equal
to the corresponding charged Higgs mass).

In Table 1 we estimate the 
enhancement ratio $R$ as a function of the technirho mass
at LEP1.3 and give predictions for the expected enhancement at LEP1.6.
For example suppose we require $R\approx 10$ in order to account
for the ALEPH data.
Then according to Table 1
such an enhancement implies that $m_{\rho_{TC}}\approx 160$ GeV
and $\Gamma_{\rho_{TC}}\approx 15$ GeV. In this example
the estimated technirho mass
is exactly equal to the CM energy of LEP1.6 which will
be sitting at the peak of the technirho resonance with
\beq
R(\sqrt{s}=160GeV)\approx 
\frac{m_{\rho_{TC}}^2}{\Gamma_{\rho_{TC}}^2}\approx 114
\eeq
corresponding to a production
cross-section of 55 GeV charged technipions of
about 63pb!
Of course the exact enhancement factor is poorly
determined at LEP1.3 so in Table 1 we have considered
technirho masses in the range 150-200 GeV. 
In all cases it is clear that LEP1.6 will always
be closer to the peak of the resonance leading to {\em much}
greater enhancements than those presently observed.
Table 1 is only meant as a
rough guide to the expected rates and if LEP1.6
does see a strong enhancement of the 4-jet rate
it will be a straightforward task to use
Eqs.\ref{width},\ref{ratio} to determine accurately the
mass and width of the technirho.

\vspace{0.25in}
  
\begin{tabular}{|c|c|c|c|} \hline
$m_{\rho_{TC} }$ (GeV) & $\Gamma_{\rho_{TC}}$ (GeV) & 
$R(\sqrt{s}=133GeV)$ & $R(\sqrt{s}=160GeV)$  \\ \hline
150 & 11 & 20 & 41\\ \hline
160 & 15 & 10 & 114\\ \hline
170 & 18 & 6 & 41 \\ \hline
180 & 21 & 4.5 & 17 \\ \hline
200 & 28 & 3 & 7 \\ \hline
\end{tabular}

{\footnotesize
Table 1: Predictions of the model for technirho masses
in the range 150-200 GeV. 
For each mass value $m_{\rho_{TC} }$ we 
tabulate the width $\Gamma_{\rho_{TC}}$
(calculated from Eq.\ref{width}) and the 
ratio $R$ (calculated from Eq.\ref{ratio})
of the charged technipion production cross-section
to the charged Higgs production cross-section for
LEP1.3 and LEP1.6 energies. Absolute values of
cross-sections may easily be obtained from $R$ by noting that 
the tree-level cross-section for 55 GeV charged Higgs production
at LEP1.3 (LEP1.6) is 0.43pb (0.55pb).} 

\vspace{0.25in}

In order to understand the above VMD results, and obtain other
predictions it is necessary to give some details of the model.
Here we shall present a stripped down version of the low scale technicolour
model introduced in ref.\cite{lstc} which is sufficient for our purposes.
In the complete model there are three ``generations'' of technifermions,
one of which is very heavy and breaks electroweak symmetry,
and two of which are light and accessible to LEP \cite{lstc}.
Here we will retain the single Higgs doublet $H$ of the standard model,
and consider only one of the two light generations of technifermions
for simplicity. The model is based on
the gauge group
\beq
SU(2)_{TC}\otimes SU(3)_C\otimes SU(2)_L\otimes U(1)_Y
\eeq
where we have added to the standard model gauge group
a new confining QCD-like gauge group $SU(2)_{TC}$ which
is asymptotically-free and confines at ${\Lambda}_{TC}=75-100 GeV$.
We assume that the usual three families and Higgs doublet 
are TC singlets, and transform in the usual way under the
standard part of the gauge group.
In the stripped down low scale technicolour model here
we shall introduce only one doublet of technifermions
$(p,m)$, which have electric charges $(1/2,-1,2)$, respectively,
and which transform in an anomaly-free way under the full
gauge group as,
\beq
\begin{array}{ccl}t_L=\left( \begin{array}{c}p_L \\ m_L \end{array}
\right) &\sim& (2,1,2,0)\\
{p_R}   &\sim& (2,1,1,1/2)\\
{m_R} &\sim& (2,1,1,-1/2)
\end{array}
\eeq
It is worth emphasising that this is not an extended TC model
since the technifermions have no additional couplings to the
ordinary quarks and leptons beyond those of the standard model.

In the present model electroweak symmetry is broken
predominantly by the usual Higgs vacuum expectation value (VEV) $v$,
since ${\Lambda}_{TC}<v$. We assume that the technifermions have 
Yukawa couplings to Higgs doublet of the form 
$\lambda_{p}\bar{p_{R}}H^{c\dagger}t_{L}$,
$\lambda_{m}\bar{m_{R}}H^{\dagger}t_{L}$ resulting in technifermion
current masses
$m_{p}=\lambda_{p}v$, $m_{m}=\lambda_{m}v$.
We expect TC condensates to form near the TC confinement scale
but now there is a vacuum alignment problem
which depends on the technifermion current masses.
The TC gauge forces tend to favour the
chiraly invariant condensate of the form
$<\bar{t^c_{L}}t_{L}+\bar{t^c_{R}}t_{R}>\neq 0$, 
while the current mass terms prefer the chiral symmetry
breaking condensates of the form,
$<\bar{t_{L}}t_{R}+\bar{t_{R}}t_{L}>\neq 0$.
\footnote{In the complete model it is likely that 
both sorts of condensates form, with the first lighter generation
of technifermions yielding chirally invariant condensates,
and the heavier second
generation yielding chiral symmetry breaking condensates.}
We assume here that the latter condensates form,
leading to global chiral symmetry breaking in the TC sector,
\beq
SU(2)_L\otimes SU(2)_R\rightarrow SU(2)_{L+R}
\eeq
yielding a triplet of technipions
$\pi_{TC}^{\pm ,0}\sim \bar{t}\sigma^{\pm ,3}\gamma_5t$, with
$<0|j^a_{\mu 5}|\pi_{TC}^b>=if_{TC}q_{\mu}\delta^{ab}$,
where $f_{TC}\sim 15-20 GeV$, and the current is
$j^a_{\mu 5}=\bar{t}\gamma_{\mu}\gamma_5\sigma^at$.

The current masses of the technifermions
break the chiral symmetry of the 
technidoublet resulting in a physical technipion mass
analagous to the way in which the physical pion mass
results from explicit quark masses.
The technipion mass $m_{\pi_{TC}}$ may be estimated by scaling up the
usual result for the ordinary pion mass $m_{\pi}$,
\beq
m_{\pi_{TC}}= m_{\pi}\sqrt{\left(\frac{m_{p} + m_{m}}{m_u + m_d}
\right)\frac{f_{TC}}{f_{\pi}}}.
\eeq
Assuming that the mass of the technipions is around 50-60 GeV
we thus deduce that $m_{p}+m_{m}\sim 10$ GeV.
The charged technipions $\pi_{TC}^{\pm}$ will decay 
predominantly via virtual W exchange, analagous to ordinary charged
pion decay. Thus for example the width into leptons is given by,
\beq
\Gamma (\pi_{TC}^{\pm}\rightarrow l^{\pm}\nu_l)
= \frac{f_{TC}^2}{f_{\pi}^2}\frac{m_l^2}{m_{\mu}^2}
\frac{m_{\pi_{TC}}}{m_{\pi}}\frac{1}{(1-\frac{m_{\mu}^2}{m_{\pi}^2})^2}
\Gamma (\pi^{\pm}\rightarrow \mu^{\pm}\nu_{\mu}).
\eeq
The largest such decay channels are thus $cs$ and $\tau \nu_{\tau}$
(the $cb$ channel is suppressed by $V_{cb}$). The $cs$ channel
is responsible for some of the 
4-jet events observed by ALEPH. Since the
decay goes via the $W$ couplings
the $\tau \nu_{\tau}$ channel should also be observed
at about 1/3 the rate of the $cs$ channel due to colour,
although there are calculational uncertainties due to 
QCD corrections (and the charm mass). In addition there may be
substantial model dependence if the simple TC model 
presented here is extended.
Nevertheless one would expect
that the $\tau \nu_{\tau}$ decay channel should 
have a significant branching fraction at some level 
and should sooner or later be observed.

In addition to the charged technipions 
$\pi_{TC}^{\pm}$ the model clearly also predicts neutral technipions
$\pi_{TC}^{0}$ of mass around 55 GeV.
The neutral technipions $\pi_{TC}^0$ 
decay via virtual Z exchange,
with the largest partial width,
\beq
\Gamma (\pi^{0}_{TC}\rightarrow \bar{b}b)
\approx
3\left(\frac{m_b^2}{m_l^2}\right)
\Gamma (\pi^{\pm}_{TC}\rightarrow l^{\pm}\nu_l).
\eeq
$\pi_{TC}^0$ will also decay into two photons via a chiral symmetry
suppressed anomalous $\pi_{TC}^0\gamma \gamma$ coupling \cite{lstc}.
Neutral technipions may be produced singly at LEP
via the production mechanism:
\beq
e^+e^-\rightarrow \gamma^{\ast},Z^{\ast}\rightarrow 
\rho_{TC}^{\ast} \rightarrow \pi_{TC}^0 {\omega}_{TC}^{\ast}
\label{mechanism2}
\eeq
The techniomega couples to charged leptons $l^{\pm}$
in the complete model \cite{lstc}
leading to signatures of the type:
$\bar{b}bl^+l^-$ or $\gamma \gamma l^+l^-$
where the mass of the $\bar{b}b$ or $\gamma \gamma$ is equal to 
that of the $\pi_{TC}^0$, and the mass of the $l^+l^-$
peaks near the techniomega mass. The precise rate is difficult to
estimate \cite{lstc}.

Apart from the low-scale technipions, the technidoublet $t=(p,m)$
will give rise to the technivector mesons $V$ mentioned above, which are
analagous to the QCD vector resonances.
For example we may expect a $J^{PC}=1^{--}$
technirho isotriplet ${\rho}_{TC}^{\pm ,0}$ and techniomega isosinglet
${\omega}_{TC}^0$ with masses in the range 150-200 GeV. 
The vector masses are given by scaling up
the ordinary $\rho$ and $\omega$ mass 
$m_{\rho_{TC} ,\omega_{TC}}\approx m_{\rho ,\omega}\frac{f_{TC}}{f_{\pi}}$.
The technidoublet $t$ has photon and Z couplings,
\beq
A_{\mu}\bar{t}\gamma^{\mu}Qt
+Z_{\mu}\bar{t}\gamma^{\mu}(\Gamma_V +\Gamma_A\gamma_5)t
\eeq
where 
\beq
Q=e\frac{\sigma^3}{2}, \ \ \Gamma_V=\frac{e}{\tan 2\theta_w}
\frac{\sigma^3}{2}, \ \
\Gamma_A=\frac{-e}{\sin 2\theta_w}\frac{\sigma^3}{2}.
\label{couplings}
\eeq
Using vector meson dominance (VMD)
arguments, combined with scaling-up
arguments, we write,
\beq
\bar{t}\gamma^{\mu}
\frac{\sigma^a}{2}t\rightarrow
\frac{m_{{\rho}_{TC}}^2}{g_{{\rho}_{TC}}}\rho^{a\mu }_{TC}
\eeq
where $g_{{\rho}_{TC}}\approx g_{\rho}$,and $g_{\rho}=\sqrt{12\pi }$.
This implies the following technirho
\footnote{Note that according to Eq.\ref{couplings}
the isosinglet techniomega has
no direct coupling to the photon or $Z$ in this model.}
couplings involving the photon and Z
\beq
\frac{m_{{\rho}_{TC}}^2}{g_{{\rho}_{TC}}}
\rho_{TC}^{0,\mu}\left[eA_{\mu} +
\frac{e}{\tan 2\theta_w}Z_{\mu}\right].
\eeq
A similar line of reasoning leads to technirho couplings to
technipions given by
\beq
- g_{{\rho}_{TC}} \epsilon_{abc}
\rho_{TC}^{0,a,\mu} \pi_{TC}^b(\partial^{\mu}\pi_{TC}^c).
\eeq
The above VMD results were in fact used to give the earlier prediction for 
$R$ in Eq.\ref{ratio}
due to the mechanism in Eq.\ref{mechanism}.

One may wonder about the contribution of charged technipions to
$b\rightarrow s\gamma$.
While charged Higgs of 55 GeV
would give a very large contribution to $b\rightarrow s\gamma$,
in this model technipions of a similar mass
do {\em not} contribute to this process.
This is because the technifermions in this model
have no direct couplings to ordinary fermions
(i.e. no extended technicolour couplings)
other than via the standard model Higgs doublet
and $W,Z$ couplings. However one would expect the charged
technipions to be produced in the top quark decays
at the Tevatron $t\rightarrow \pi_{TC}b$ at some level.

Also one may be concerned about 
the constraints on technicolour from high precision
tests of the standard model. 
At LEP1 energies the technirho will contribute to
the oblique corrections to the photon and Z propagators. 
For example the contribution to the S parameter \cite{7} for
a single technidoublet is naively estimated to be
$S\approx 0.3\left(\frac{2}{3}\right) \approx 0.2$.
\footnote{Note that
the TC contribution depends on
scale independent ratios like
$\frac{f_{TC}}{m_{{\rho}_{TC}}}$ 
so is independent of the fact that the technicolour scale 
(and the technirho mass) is low, assuming that the technirho
is much heavier than the $Z$ boson. 
However our estimate relies on VMD and
large $N_{TC}$ scaling arguments in order to estimate the contribution
to dispersion relations, and such may be unreliable 
in TC theories \cite{lane}. }
For two technidoublets our naive estimate
satisfies the phenomenological constraint \cite{langacker}
$S<0.38(0.46)$ at $90\%(95\%)$ CL,
but this estimate is likely to be unreliable since the first generation
technidoublet is not assumed to form QCD-like condensates. 
For three technidoublets
we must begin to appeal to such unreliability of the estimates.

Finally it should be pointed out that there are other models
in the literature which also have a low TC scale, for example 
ref.\cite{EL}. These authors were concerned with signals at 
hadron colliders and pointed out that the technirho
could be produced at the Tevatron and could
decay into a technipion $\pi^{\pm}_{TC}$ 
plus a longitudinally polarised $W_L$,
yielding a characteristic signature.
Although such decays are suppressed 
relative to the two technipion decays
by a factor $(f_{TC}/245 GeV)^2$, they would
dominate if the technipions are heavier  
than half the technirho mass. However this condition is clearly
{\em not} fulfilled here since the technipions
have masses around 55 GeV while the technirho has a mass
in the range 150-200 GeV. Therefore 
such decays will have a very small branching fraction at the
one per cent level and are unlikely to be 
observable at the Tevatron, although
interestingly such decays may be observed in the high statistics clean 
environment of LEP1.6.

To summarise, we have proposed that 
approximately a half of the anomalous 4-jet events seen
by ALEPH at LEP1.3 are due to resonantly enhanced charged technipion
production, involving charged technipions. 
If this interpretation is correct then we should
expect even more enhancement in the rate at LEP1.6 since 
the CM energy will be closer to the pole of the technirho, 
as is clearly seen in Table 1. 
The discovery of a light technirho
would be the biggest bombshell in high-energy physics since
the discovery of the $J/\psi$, and we look forward to the 
next LEP runs in June with some excitement.

{\bf Acknowledgement}

I would like to thank A. Vayaki and P. Janot
for useful communications.

\bibliographystyle{unsrt}

\end{document}